\documentclass[12pt,preprint]{aastex}

\newcommand{\beq}{\begin{equation}}
\newcommand{\eeq}{\end{equation}}
\def\la{\hbox{\raise.35ex\rlap{$<$}\lower.6ex\hbox{$\sim$}\ }}
\def\ga{\hbox{\raise.35ex\rlap{$>$}\lower.6ex\hbox{$\sim$}\ }}

\def\beq{\begin{equation}}
\def\eeq{\end{equation}}
\def\beqa{\begin{eqnarray}}
\def\eeqa{\end{eqnarray}}
\def\bseq{\begin{subequations}}
\def\eseq{\end{subequations}}

\newcommand\lsim{\mathrel{\rlap{\lower4pt\hbox{\hskip1pt$\sim$}}
        \raise1pt\hbox{$<$}}}
\newcommand\gsim{\mathrel{\rlap{\lower4pt\hbox{\hskip1pt$\sim$}}
        \raise1pt\hbox{$>$}}}

\begin{document}

\title{Probing {Gravity} with Spacetime Sirens} \author{C\'edric
Deffayet\altaffilmark{1} \& Kristen Menou\altaffilmark{2}}
\altaffiltext{1}{APC UMR 7164 (CNRS, Univ. Paris 7, CEA, Obs. de
Paris) \& GReCO/IAP UMR 7095 (CNRS, Univ. Paris 6) Paris, France}
\altaffiltext{2}{Department of Astronomy, Columbia University, New
York, NY 10027, USA}

\begin{abstract}
A gravitational observatory such as {\it LISA} will detect coalescing
pairs of massive black holes, accurately measure their luminosity
distance and help identify a host galaxy or an electromagnetic
counterpart. If dark energy is a manifestation of modified gravity on
large scales, gravitational waves from cosmologically-distant
spacetime sirens are direct probes of this new physics.  For example,
a gravitational Hubble diagram based on black hole pair luminosity
distances and host galaxy redshifts could reveal a large distance 
extra-dimensional leakage of gravity. Various additional signatures
may be expected in a gravitational signal propagated over cosmological
scales.
\end{abstract}

\section{Introduction}

Evidence for an accelerating expansion of the Universe is getting
stronger. A measurable dimming of distant type Ia supernovae
\citep[e.g.,][]{ast06,riess07,woo07}, CMB data
\citep[e.g.,][]{spe06} and additional cosmological probes
\citep[e.g.,][]{wri07} indicate that a repulsive "dark energy"
comprises 73\% of the Universe's content.  Characterizing the
properties of dark energy and uncovering its physical nature are two
of the most important goals of modern cosmology.  Current
observational strategies include tests of the possibility that dark
energy arises from a failure of general relativity on cosmological
scales \citep{alb06}.

Essentially all astronomical measurements are performed via
electromagnetic waves. The availability of accurate gravitational wave
measurements, within the next decade or so, will thus be a significant
development. In particular, since the propagation of photons and
gravitons {could} differ at a fundamental level, we argue here that
gravitational waves emitted by cosmologically-distant ``space-time
sirens,'' such as coalescing pairs of massive black holes, may be used
as valuable alternative probes of dark energy physics.

Black holes with masses $\gsim 10^6 M_\odot$ are present at the center
of numerous nearby galaxies \citep[e.g.][]{kr95,mag98}. As such
galaxies collide over cosmic times, their central black holes
coalesce, releasing $\gsim 10^{58}$~ergs of binding energy in the form
of gravitational waves (hereafter GWs). To measure the GWs emitted by
these cosmologically-distant space-time sirens, ESA and NASA will
build the Laser Interferometer Space Antenna, LISA\footnote{{\tt
http://lisa.nasa.gov/}}.

\section{Gravitational Measurements}

GWs emitted by black hole binaries have the unusual property of
providing a direct measure of the luminosity distance, $D_L$, to the
black holes, without extrinsic calibration. Owing to the highly
coherent nature of GW emission \citep{schu86}, the amplitude (or
strain), $h_{+\times}$, frequency, $f$, and frequency derivative,
$\dot f$, of the leading order (quadrupolar) GW inspiral signal scale
as
\begin{eqnarray} 
h_{+\times} (t) & \propto & \frac{\left[ (1+z) M_c \right]^{5/3}
f^{2/3}}{D_L}, \\ \dot f (t)& \propto & \left[ (1+z) M_c \right]^{5/3}
f^{11/3},
\end{eqnarray}
 where $+\times$ represents the two transverse GW polarizations, $M_c=
(m_1 m_2)^{3/5} / (m_1+m_2)^{1/5}$ is the black hole pair ``chirp''
mass and $z$ its redshift. Provided the GW source can be reasonably
well localized on the sky, an extended observation of the chirping
signal leads to precise measurements of $h_{+\times}$, $f$, $\dot f$
and thus $D_L$, independently. LISA's orbital configuration allows for
a ``triangulation'' of GW sources on the sky to within a solid angle
$\delta \Omega \sim 1$~deg$^2$ \citep{cut98,vec04}, thus providing
very accurate distance measurements, with $\delta D_L / D_L < 1\%$ at
$z \lsim 2$ typically \citep{cut98,hug02,vec04,lh06}.

Recently, the possibility of identifying the individual host galaxy in
which a pair of merging black holes seen by LISA is to be found has
been explored in some detail \citep{hohu05,koc06,koc07}. Prospects for
such identifications at redshifts $z \lsim 3$ are good and
identifications out to $z \sim 5$-$7$ may even be possible in some
cases \citep{koc07}. A unique host galaxy identification can be
achieved through coordinated observations with traditional telescopes,
either to survey the LISA-triangulated area for unusual galactic
properties or central activity after the coalescence, or by monitoring
in real time the sky area for unusual electromagnetic emission, as the
coalescence proceeds. Typically $\gsim 10^{53}$~ergs of kinetic energy
are delivered to the recoiling black hole remnant \citep[e.g.,][and
references therein]{scbu07} and its environment. The disturbed gas
surrounding coalescing black holes could thus power bright
electromagnetic emission during and after coalescence
\citep{armnat02,milphi05,dot06}, permitting the coincident
identification of a unique host galaxy.

\section{Gravitational Hubble Diagram}

A consequence of successfully identifying the host galaxies of
coalescing black hole pairs is the possibility to draw a gravitational
Hubble diagram, i.e. one that relates the gravitational luminosity
distances, $D_L$, of space-time sirens to the electromagnetic
redshifts, $z$, of their host galaxies.

The solid line in Fig.~\ref{fig:one} shows a model fit to a Hubble
diagram from a recent type Ia supernova sample with a dark energy
equation of state $P/\rho \equiv w=-1.05 \pm 0.3$ (at $3 \sigma$,
shown by the shaded region). Luminosity distances are expressed in
terms of the usual distance modulus, $\mu = 5 \log_{10}(D_L/\,1\,{\rm
Mpc}) +25$.  The individual supernova data points of the gold sample
of \cite{riess07} are also shown. A flat universe with $H_0 =
72$~km~s$^{-1}$~Mpc$^{-1}$ and a matter density $\Omega_m=0.27$ is
assumed.

The general interest of a gravitational Hubble diagram can be
illustrated by considering alternative explanations for the apparent
dimming of distant type Ia supernovae.  If this dimming were caused by
dust attenuation, photon-axion conversions or an intrinsic evolution
of the supernova population \citep[e.g.,][]{Csaki,mir06,duax}, rather than
stretched distances, this would become apparent in comparisons between
gravitational and electromagnetic Hubble diagrams. Indeed, in each of
these alternative dimming scenarios, photon- and graviton-based Hubble
diagrams would be fundamentally discrepant since self-calibrated
gravitational distance measurements are not susceptible to any
significant bias from absorption, scattering, reddening, or
axion-conversion.

In practice, however, the value of such a comparison is limited by
line-of-sight matter inhomogeneities, which generate ``weak lensing''
uncertainties on any photon or graviton $D_L$ measurement
\citep{hohu05,koc06,dal06}.  While the lensing effect can be averaged
out over the many random lines-of-sights available with the supernova
data (i.e., data points in Fig.~\ref{fig:one}), it {may} not be the
case for coalescing black hole pairs if LISA merger event rates are
modest \citep[e.g., a few tens per year at $z \lsim
5$;][]{men01,wl03,ses04,mic07}. Weak lensing errors on individual
measurements amount to distance uncertainties ranging from $\delta D_L
/ D_L \simeq 1\%$ at $z=0.5$ to $\delta D_L / D_L \simeq 10\%$ at
$z=5$ \citep[e.g.][]{koc06}. The corresponding $3 \sigma$ distance
modulus uncertainties exceed the $3 \sigma$ confidence contours on the
current dark energy model fit (shaded region in Fig.~\ref{fig:one}) at
$z \gsim 0.7$, which makes comparisons between photon- and
graviton-based Hubble diagrams imprecise even at moderate
redshifts. The extent to which LISA events can be used to perform
meaningful comparisons with photon-based Hubble diagrams will thus
depend strongly on the actual distribution of massive black hole
merger events with redshifts and the corresponding efficiency of host
galaxy identifications.

\section{Modified Gravity}

The greatest prospect for dark energy science with gravitational waves
may lie in exploring new physics on cosmological scales. The
possibility that accelerated expansion results from a failure of
general relativity has fueled much theoretical work on large scale
modifications of gravity over the past few years.  Since building a
satisfactory theory of modified relativistic gravity is a formidable
task, any insight that can be gained from direct observational
constraints on the linearized gravitational wave regime cannot be
overlooked.

One {may expect} gravity modifications to contain a new length scale,
let us call it $R_c$, beyond which gravity deviates from general
relativity. In order to explain the observed accelerated expansion of
the Universe, this scale is expected to be of the order of the current
Hubble radius $H_0^{-1}$. Modified gravity must also pass standard
tests of general relativity on scales much shorter than $R_c$, e.g. in
the solar system and in the strong field regime of binary pulsars. An
existence proof of modifications of this type is given by DGP gravity
\citep{dgp}, a braneworld model with an infinite (possibly many) extra
dimension. In this model, which also leads to an accelerated expansion
of the Universe \citep[]{def01,def02}, gravity is intrinsically higher
dimensional\footnote{\citet{gor06}, \citet{char06}, \citet{def06},
\citet{dva06} \citet{izu07} and \citet{gre07} discuss the stability of
the DGP self-accelerating phase.}. To leading order, the gravitational
potential has the standard $1/D$ behavior at distances $D$ smaller
than $R_c$, while it behaves 5-dimensionally at larger distances
($1/D^2$). Moreover, the stronger the gravitational field, the closer
the theory is to general relativity. As a consequence the model passes
the PPN tests in the solar system, with possible deviations emerging
in upcoming generations of solar system measurements
\citep{lue03,dva03}. The GW emission of classical astrophysical
sources is also expected to closely match that of general
relativity. It is only at very large distances and low curvature that
gravity is beginning to "leak" in the extra dimension.

A previously unexplored consequence of extra-dimensional leakage is
that cosmologically-distant GW sources would appear dimmer than they
truly are, from the loss of GW energy flux to the bulk. Inspired by
this idea, we investigate here the consequences of this possible
modification of gravity, borrowing from the DGP model the notion that
strong field gravity (and hence GW emission) asymptotes to general
relativity, while deviations appear at very large, typically
cosmological distances, in the weak field regime. We will not deal
here specifically with GWs in the DGP model, which is the subject of a
separate study, rather, we illustrate more generally how a
gravitational observatory such as LISA may reveal cosmological
deviations in the weak field graviton propagator.

In the presence of large distance leakage {(say at distances much
larger than $R_c$)}, flux conservation over a source-centered
hypersphere requires that the GW amplitude scales with distance
$D$ from the source as
\begin{equation} \label{scaling}
h_{+\times} \propto D^{-(dim-2)/2},
\end{equation}
where $dim$ is the total number of space-time dimensions accessible to
gravity modes. Thus, for $dim \geq 5$, it deviates from the usual $
h_{+\times}(D) \propto 1 /D$ scaling. The scaling in
Eq.~(\ref{scaling}) is consistent with explicit GW calculations in
spacetimes with compact extra-dimensions \citep{card03,barsol03} and
also applies to models where extra dimensions open up only at large
distances \citep{grs,dgp,dgs}.

\begin{figure}[ht!]
\begin{center}
\plotone{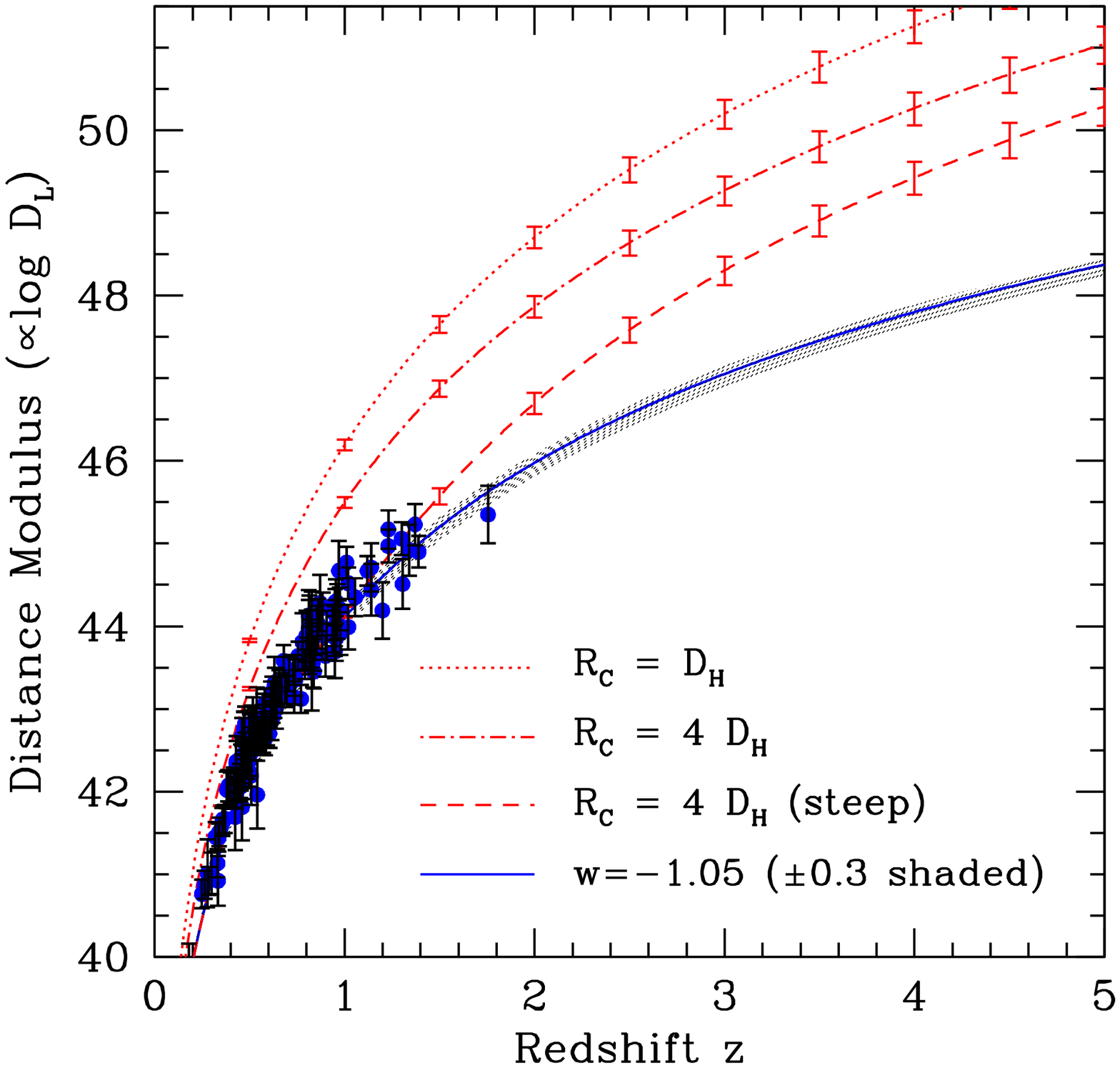}
\caption{Electromagnetic Hubble diagram of a standard dark
energy model (solid line + shaded region) and hypothetical
gravitational Hubble diagrams for three {modified gravity} scenarios
with biased gravitational luminosity distances from {large distance
gravitational} leakage, beyond a cross-over scale $R_c$ (top three
lines).\label{fig:one}}
\end{center}
\end{figure}

The top three lines in Fig.~\ref{fig:one} show gravitational Hubble
diagrams, i.e. the expected locus of $D_L$ and $z$ measurements from
black hole pairs and host galaxies, in three simple scenarios with
$dim = 5$ and leakage beyond a scale $R_c$ obeying
Eq.~(\ref{scaling}). To allow arbitrary possibilities for the
transition at the cross-over scale, we adopt $h_{+\times}(D_L) \propto
(D_L [1+(D_L/ R_c)^{n/2}]^{1/n})^{-1}$, where $n$ determines the
transition steepness. In the three cases shown, cross-over scales of a
few Hubble distances, $R_c = 1$-$4~D_H \sim 2$-$9$~Gpc were used and
values $n=1$ or $10$ (``steep'') for the transition steepness were
adopted.  Error bars show the magnitude of $1 \sigma$ weak lensing
uncertainties on $D_L$ measurements due to line-of-sight matter
inhomogeneities.

The electromagnetic Hubble diagrams of each scenario {is assumed} to
mimic the cosmology of a $w=-1$ dark energy model (solid line),
as expected if these scenarios produce an accelerated background
cosmological FLRW space-time which affects the geodesic motion of
photons in the standard way.  As a result of the different
propagation of GWs and large distance leakage of GW energy flux,
however, measured GW amplitudes, $h_{+\times}$, are reduced and
gravitational $D_L$ values are overestimated (Eqs.~[1]--[2]) for
sources beyond the cross-over scale in these scenarios. The
corresponding graviton Hubble diagrams would thus deviate from the
electromagnetic version (solid line) since all photon-based
measurements are standard, by construction.  The discrepancy
between the graviton-measured and photon-inferred $D_L$ values could
be of order unity even at moderate redshifts. Black hole merger events
and associated host galaxies may thus reveal the {leakage of gravity}
with cross-over scales $R_c \sim $ a few Hubble distances.

Diagnostics based exclusively on electromagnetic measurements cannot
reveal gravitational leakage directly, although they may do so
indirectly \citep[e.g.,][]{lue04,wang07,zhang07}. Importantly, the
leakage may be equally difficult to identify with only a GW signal. As
is well known, all the mass combinations that determine the GW signal,
e.g. in post-Newtonian expansions of the inspiral signal, are
redshifted by a factor $1+z$ and thus degenerate with redshift
\citep[e.g.,][]{hug02}. Provided that general relativistic black holes
and GW emission are accurately recovered on sub-cosmological scales in
 the modified gravity scenarios considered, leakage on
cosmological scales would lead to a hidden bias on the redshift
(overestimated) and black hole masses (underestimated), as inferred
from adopting a background cosmology that ignores leakage
altogether. Only if the redshift of the host galaxy were to become
available would the discrepancy with the gravitational measurement
become apparent, as illustrated in Fig.~\ref{fig:one}.  Paradoxically,
this could limit our abilities to identify the host galaxy of a
merging black hole pair, by invalidating search strategies that rely
on a (biased) value of the gravitational luminosity distance to define
redshift cuts on potential host galaxies \citep{koc06}.

{We note} that evidence of GW leakage could also emerge in different
contexts, e.g. from low redshift short gamma-ray bursts detected by a
network of ground-based GW detectors \citep{dal06}, or from an
unexpectedly low amplitude of the present-day GW background relative
to its CMB-normalized value (A. Buonanno, priv. communication).

Despite the strong GW leakage suggested by the above energetic
arguments, one should be very cautious in evaluating the significance
of the deviations shown in Fig.~\ref{fig:one}. Indeed,
Eq.~(\ref{scaling}) says very little about the cross-over transition
physics and other details of GW propagation on cosmological scales.
In particular, even though the scaling given in Eq.~(\ref{scaling})
does apply to the DGP model, the "infrared transparency" effect
\citep{dgs} can be shown to result in a considerable increase of the
distance at which GW leakage is manifested in DGP, reaching scales
much beyond $R_c$ for sources with frequencies relevant to LISA
\citep{dmz}. Therefore, the DGP model
does not necessarily produce significant deviations between
gravitational and electromagnetic luminosity distances and it would,
in fact, be difficult to distinguish from general relativity on the
basis of the test illustrated in Fig.~\ref{fig:one}.  On the other
hand, since it is presently {unclear} whether infrared transparency
and its consequences for GW propagation are generic to all cases of
higher-dimensional gravity with large distance leakage {(in particular
when Lorentz symmetry is broken)}, Fig.~\ref{fig:one} remains useful
in providing a good measure of potential GW leakage.

\section{Additional Signatures}

Large distance leakage is only one of several possible modified
gravity signatures in the GW signal from cosmologically-distant
spacetime sirens. We discuss a few additional possibilities here,
voluntarily adopting a simple phenomenological approach.

A first class of signatures resides in the GW polarization signal.  In
many modified gravity scenarios, additional polarizations exist beyond
the two transverse quadrupolar ($+\times$) modes of general relativity
\citep[e.g.,][]{will06}. This is the case, for instance, in
scalar-tensor theories \citep[including the f(R)
variety;][]{frscal,magnic00, nak01} and vector-tensor theories
\citep{bek04,jac04}.  Braneworld gravity scenarios {are no exception
since the Kaluza-Klein graviton extra polarizations can also in
principle be radiated by {various types of} sources}.  The no-hair
theorem is traditionally invoked to forbid any extra polarization
modes in the signal from two coalescing black holes. However, it is
well known that black hole unicity theorems fail to apply in the usual
sense to higher dimensional theories \citep{er}. Moreover, in DGP
gravity the graviton extra polarization is expected to show up at
large distances even for (possibly static) spherically symmetric
space-times \citep{def02b}. Black holes can also be hairy in theories
of massive gravity (where the graviton carries extra polarizations
with respect to those of a massless graviton), in particular when
Lorentz invariance is broken  \citep[][and references
therein]{blas06,blas07,dtz07}.

A second class of signatures is related to the GW signal propagation
velocity which, in modified gravity scenarios, can differ from the
speed of light. Propagation can be subluminal \citep[e.g.,][]{dtt05}
or superluminal \citep[e.g.,][]{jac04}. The possibility to time a GW
signal propagated over cosmological distances, relative to the signal
from a prompt electromagnetic counterpart causally associated with the
black hole merger, may thus offer additional diagnostics of
large-scale modified gravity. In addition, signatures may emerge
because the number of spacetime dimensions available to gravity is
odd. In this case, even a massless mode has a Green's function which
does not vanish inside its light cone \citep[see
e.g.,][]{card03}. Consequences for GW propagation in braneworld
gravity have been explored in part \citep{barsol03}, but not fully in
the case when gravity is modified at large distances. Thus, the
oddness of spacetime on cosmological scales, like in DGP gravity,
could add to the set of modified gravity signatures.

Finally, a third class of signatures relates to the phase of the GW
signal, which could deviate from general relativistic expectations
once propagated over cosmological distances.

In summary, GWs from cosmologically-distant spacetime sirens may be
valuable alternative probes of {modified gravity} since various
signatures may exist in a GW signal propagated over cosmological
distances. For some of these signatures to become apparent, the
identification of a host galaxy or an electromagnetic counterpart to
the spacetime siren is required. Conversely, the absence of such
signatures would constrain gravity modifications and provide a
consistency check on other methods employed to discover the nature of
dark energy.

\acknowledgements
This work has made use of the advanced cosmology calculator
\citep{wri06}. It is a pleasure to thank A. Buonanno,
{G. Esposito-Farese}, G. Gabadadze, Z. Haiman, D. Helfand and
B. Kocsis for useful inputs and discussions.

\clearpage 

\end{document}